# Multi-material heterogeneous integration on a 3-D Photonic-CMOS platform

**Luigi Ranno**[1,†] **, Jia Xu Brian Sia**[1,2,†]**, Khoi Phuong Dao**[1] **and Juejun Hu**[1,*]

[1]Department of Materials Science & Engineering, Massachusetts Institute of Technology, Cambridge, M.A., USA.

[2] Centre for Micro- & Nano-Electronics (CMNE), Nanyang Technological University, Singapore.

[†] These authors contributed equally to this work.

*hujuejun@mit.edu

# Abstract

Photonics has been one of the primary beneficiaries of advanced silicon manufacturing. By leveraging on mature complementary metal-oxide-semiconductor (CMOS) process nodes, unprecedented device uniformities and scalability have been achieved at low costs. However, some functionalities, such as optical memory, Pockels modulation, and magnetooptical activity, are challenging or impossible to acquire on group-IV materials alone. Heterogeneous integration promises to expand the range of capabilities within silicon photonics. Existing heterogeneous integration protocols are nonetheless not compatible with active silicon processes offered at most photonic foundries. In this work, we propose a novel heterogeneous integration platform that will enable wafer-scale, multi-material integration with active silicon-based photonics, requiring zero-change to existing foundry process. Furthermore, the platform will also pave the way to a class of high-performance devices. We propose a grating coupler design with peak coupling efficiency reaching 93%, an antenna with peak diffraction efficiency in excess of 97%, and a broadband adiabatic polarization rotator with conversion efficiency exceeding 99%.

# Introduction

The viability of photonics with established CMOS manufacturing capabilities enables economies of scale to be exploited, resulting in the large-scale, high-volume production of cutting edge photonic integrated circuits (PICs) based on silicon[1]. The field has advanced greatly since its inception in the early 1990s, making great strides in key applications such as optical communications[2–4], spectroscopy[5–7], and light detection and ranging (LiDAR)[8,9], to name a few. It is generally accepted that silicon photonics will have widespread impact on multiple distinct application spheres.

The standard silicon photonics platform, however, has several intrinsic limitations. Just to name a few examples, the centrosymmetric structure of silicon lacks second order $\chi^{(2)}$ nonlinearity and Pockels effects for electro-optic modulation[10]. Magneto-optical (MO) materials which represents the cornerstone of optical isolators[11,12] are not available within contemporary silicon photonic foundries. Light detection, which relies on germanium waveguide photodetectors, typically does not operate beyond L-band, even though the silicon-on-insulator (SOI) platform supports low-loss transmission from 1.1 to 3.8 μm[13–15] which is useful for spectroscopic sensing[5,7]. These limitations have motivated heterogeneous integration (HI) of various materials with silicon/silicon nitride waveguide platforms[16–19], empowering new functionalities such as on-chip sources[20–25], nonvolatile memories[17–19], nonreciprocity[11,12], and MIR photodetection[16].

Since it is necessary to bring the materials to be integrated into close proximity with the waveguide core to enhance light-material interactions, these examples of HI are performed either on devices without the standard back-end-of-the-line (BEOL) stack, or inside windows or trenches etched into the BEOL layers. Both approaches deprive the devices of access to the BEOL stack, which includes multiple metal interconnect layers and one or more SiN waveguide layers in most photonic foundry processes.[1,26,27] This is a major missed opportunity in current HI schemes, as we will show later how the BEOL layer can be harnessed to create photonic devices with unconventional functionalities or unprecedented performance. Furthermore, the latter approach necessitates a custom window opening step, incurring extra cost and processing time while elevating the risk of contamination and damage to the exposed waveguide cores. In the case of deposited materials, uniform film deposition into the trench can be challenging especially when the window etched into the backend dielectrics has a small area or a high aspect ratio.

In this work, we propose a novel HI platform, **Su**bstrate-inverted **M**ulti-**M**aterial **I**ntegration **T**echnology (SuMMIT), to address the aforementioned limitations of existing HI schemes while additionally offering critical advantages in scalability, performance, manufacturability, and packaging. We will start with introducing the basic architecture and process flow of SuMMIT, then proceed to detailing its key benefits, and finally demonstrate a cohort of high-performance silicon photonic components enabled by the platform.

**Introduction to the SuMMIT Platform**

Figure 1 illustrates the SuMMIT platform architecture and Fig. 2 depicts the process flow of implementing the platform for HI. The fabrication process starts with three separate wafers: a silicon photonics (active PIC) wafer, a silicon or glass wafer with embedded through silicon/glass vias (TSV/TGV) and (optionally) one or more redistribution layers (RDLs), and a CMOS electronics wafer, all of which are manufactured in industry standard semiconductor foundries. The PIC wafer typically consists of doped silicon and germanium, one or more SiN waveguide layers, and metal interconnections sequentially stacked on a SOI substrate. As a convention, here we use 'frontside' and 'backside' to refer to the PIC wafer top surface (with metal pads) and the bottom side of the starting SOI wafer, respectively. Next, direct-bond interconnects (DBI) are formed on both sides of the TSV/TGV wafer as well as the electrical contact pads of the PIC and CMOS wafers. The DBI surfaces are then planarized and polished (Fig. 2b) to facilitate wafer bonding. The wafers are subsequently bonded via hybrid bonding (Fig. 2c). The hybrid bonding process uses low bonding temperature to prevent damage to the BEOL metal interconnects, and it also produces a mechanically strong bonding interfaces across both dielectrics and metals to prevent delamination or deformation due to built-in stress in the PIC layers after substrate removal. After the bonding steps, the PIC wafer is inverted and the Si handler substrate is removed by deep reactive ion etching (DRIE), followed by thinning of the buried oxide (BOX) layer through chemical mechanical polishing (CMP). The last 100 nm of BOX is removed using a timed HF etch (Fig. 2d). The substrate removal process results in a smooth, topology-free surface ideally suited for subsequent processing steps, such as lithography, film deposition, die bonding, or membrane/layer transfer. This enables the universal heterogeneous integration of various photonic materials on the PIC wafer backside within a wafer-level, CMOS-backend-compatible framework (Fig. 2e).

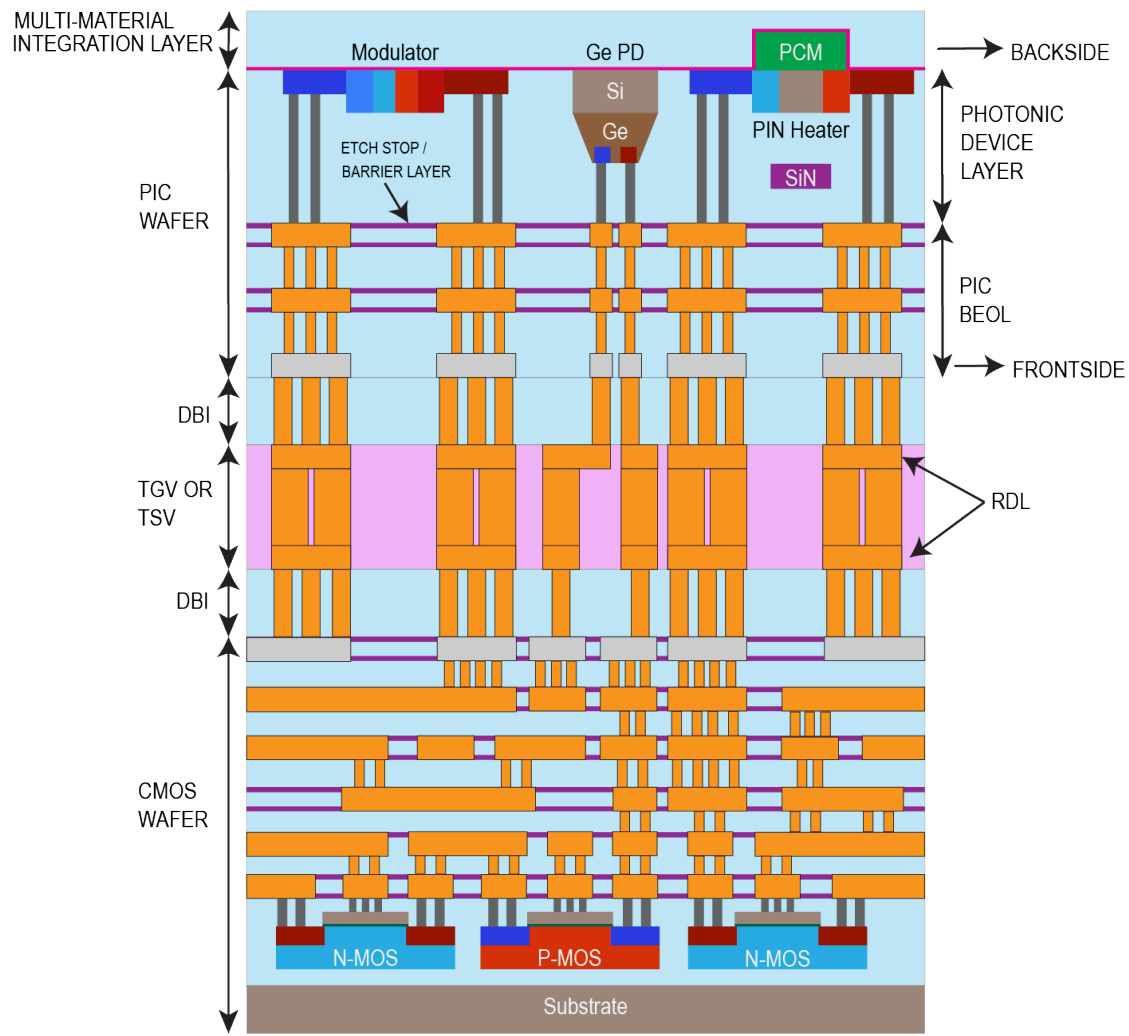

**Figure 1. Schematic cross-sectional diagram showing the SuMMIT platform**. The platform in general consists of 4 technology layers: a CMOS wafer, a TSV/TGV wafer, silicon

photonics (PIC, whose wafer substrate is removed), and a multi-material integration layer. The multi-material integration layer refers to photonic materials that are heterogeneously integrated to extend the portfolio of functionalities available to the current silicon-based photonic platform (e.g., phase change material, PCM). The entire stack can be manufactured at wafer-scale, and fabrication of the three wafers (CMOS, TSV/TGV, and PIC) requires zero change to standard foundry processes.

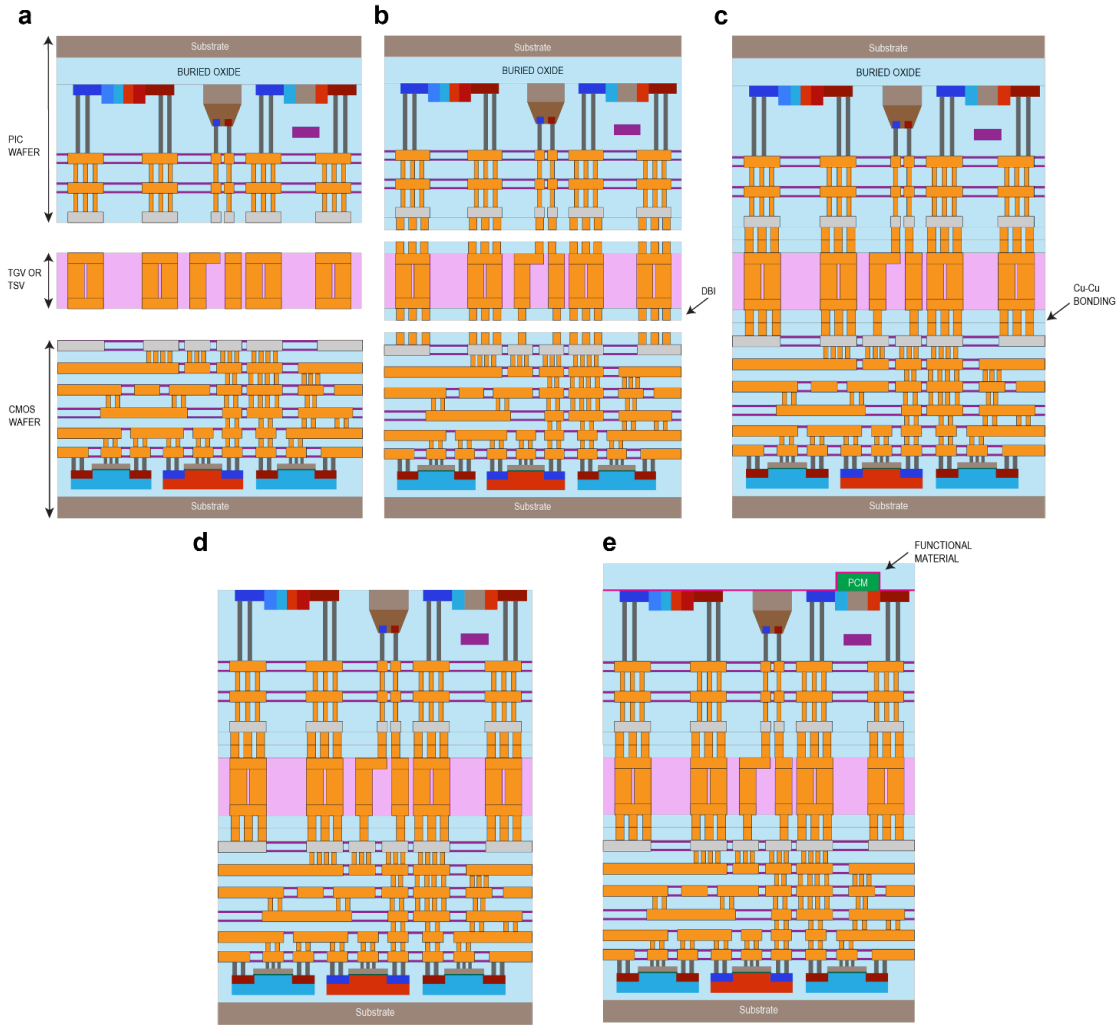

**Figure 2. Fabrication process flow of the SuMMIT platform**. **a,** Active Si PIC wafer, CMOS wafer and TSV/TGV wafer. **b,** Formation of DBIs on both sides of the TSV/TGV wafer, and on metal contact pads of the PIC and CMOS wafers. **c,** Hybrid bonding of the PIC wafer and the CMOS wafer on both sides of the TSV/TGV wafer. **d,** Silicon substrate and BOX removal of the PIC wafer. **e,** HI of beyond-CMOS functional materials at the backside of the PIC (e.g., PCMs).

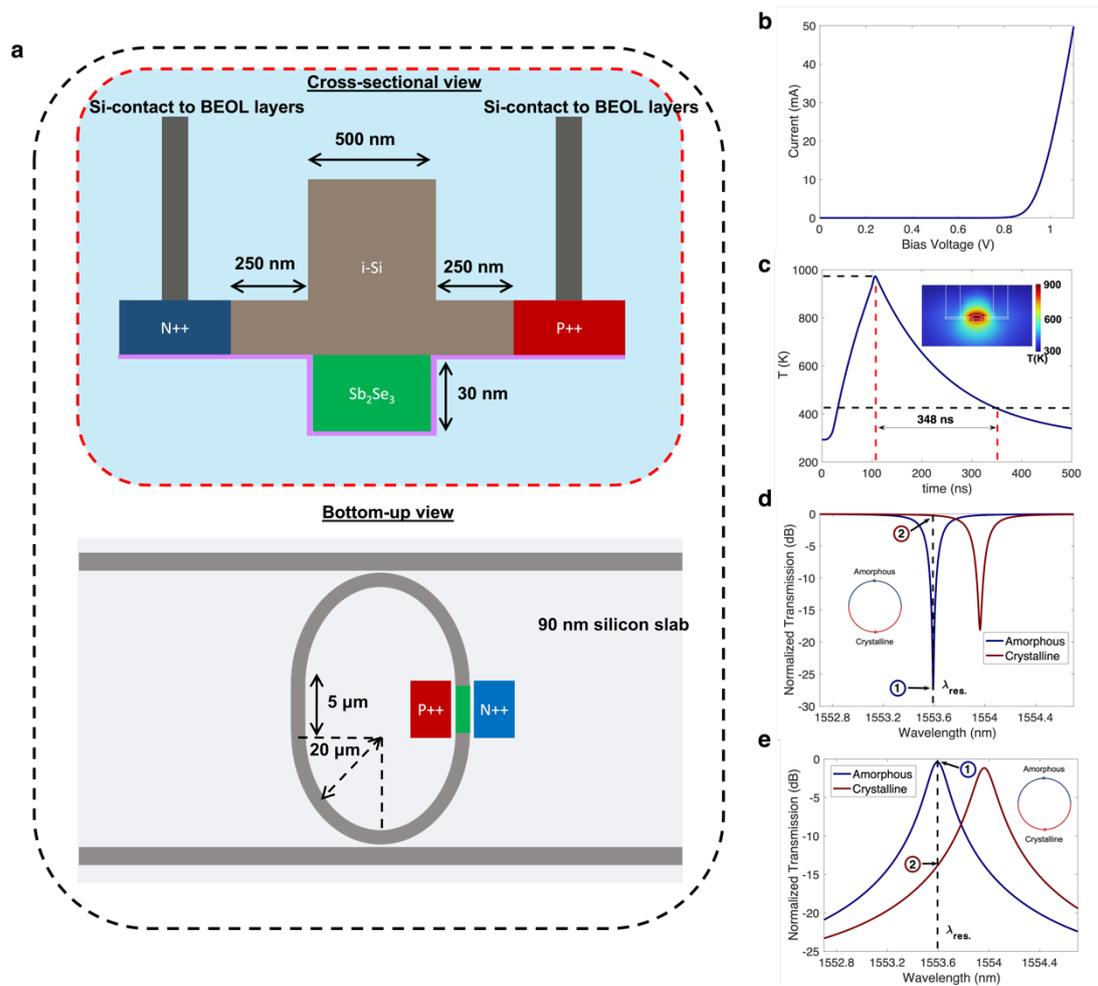

**Figure 3. HI of Sb$_2$Se$_3$ PCM on a P-I-N junction embedded in an SOI racetrack resonator, implemented on the SuMMIT platform**. **a,** Cross-sectional and bottom-up views of the P-I-N junction where Sb$_2$Se$_3$ film is heterogeneously integrated. **b,** Experimentally measured diode response in the forward bias that can be implemented for nonvolatile switching of PCMs. **c,** Transient response of the PCM-integrated P-I-N heater driven by an amorphization electrical pulse of 6 V and 100 ns duration. The inset shows a snapshot of the peak temperature profile when the amorphization pulse is applied across the junction. Both charts are simulated by the finite element method. **d-e,** Simulated **d** through and **e** drop port optical spectra of a PCM-integrated racetrack resonator.

As an example of HI using the SuMMIT platform, Fig. 3a shows the integration of a PCM, Sb$_2$Se$_3$, with a racetrack resonator to realize a nonvolatile 2 × 2 switch. The device designs discussed herein are all compliant with an active silicon PIC process reported by Sia *et al.*[1], which is commercially available through CompoundTek. We note that the general design principles illustrated here can be equally adapted to other photonic foundry processes. The Sb$_2$Se$_3$ film is 30 nm thick and 5 micron long, and is deposited and patterned to cover one of the straight sections of the racetrack (Fig. 3a, bottom-up view), where a P-I-N junction is embedded. The bending radius and the straight section of the racetrack are 20 μm and 5 μm respectively. The cross-sectional dimensions of the P-I-N junction are indicated in Fig. 3a. The measured diode characteristics of the P-I-N junction in the forward bias regime are plotted in Fig. 3b. Using the I-V data, we simulated the electrothermal behavior of the P-I-N heater. Its transient response shown in Fig. 3c implies a thermal time constant (1/e of peak temperature) of approximately 348 ns which is sufficient for melt quenching of $Sb_2Se_3$ to bypass crystallization[28,29]. The inset of Fig. 3c presents a snapshot of the simulated peak temperature profile pertaining to the P-I-N junction during amorphization, where temperatures above the melting point of $Sb_2Se_3$ (884 K) can be attained throughout the PCM volume to ensure complete amorphization. As shown in the through (Fig. 3d) and drop spectra (Fig. 3e) of the racetrack resonator simulated using the S-matrix method, electrothermal switching of the PCM between crystalline and amorphous states triggers optical switching from through to drop ports with a large contrast. The slight reduction of extinction ratio at the crystalline state compared to the amorphous state is attributed to the higher absorption of crystalline $Sb_2Se_3$[17,19]. By fixing a laser wavelength at the resonant wavelength indicated by $\lambda_{res.}$, nonvolatile transmission (states 1 and 2), and as such, electrically addressed optical memory, can be demonstrated on the platform.

The architecture and integration process described herein differs from prior wafer backside integration approaches[30,31] in that the optical and electrical I/Os are separately routed from top and bottom surfaces of the PIC to allow a much higher degree of integration, and that universal heterogeneous integration of various new materials is envisaged. The SuMMIT platform further uniquely enables a cohort of novel device designs and some examples are detailed in the following.

**Ultra-Efficient Grating Couplers**

Silicon photonics is a high-index contrast platform commensurate with dense photonic integration on-chip. However, the strong mode confinement also results in a large mode size difference between a single-mode fiber (SMF) and silicon waveguides. As a result, immense efforts have been dedicated towards developing highly-efficient optical I/O interfaces[32,33]. Among the coupling schemes, silicon grating couplers are provided within typical foundry Process Design Kits (PDKs). However, limited coupling efficiencies of approximately 50% or less are routinely quoted for foundry-compatible grating couplers[34]. The performance of grating couplers is primarily dictated by both grating directionality and mode mismatch between the diffracted pattern and the fiber mode. High grating directionality can be achieved via engineering the phase of scattered light using a corrugated pattern[35] with poly-Si overlay, which however is not accessible in most photonic foundry processes. The addition of a bottom reflector is a much simpler yet effective alternative, although it is often challenging to integrate a mirror reflector in a scalable fashion within a CMOS-compatible framework[36,37]. Furthermore, as abovementioned, within current photonic technologies, it is of note that electrical and optical I/O (grating couplers) are coupled from the chip frontside, which would impose a limit on integration densities. Leveraging on the SuMMIT architecture, we propose a highly-efficient grating coupler. Light is coupled through a SMF (angled at $\theta = 10°$ from surface normal) from the PIC backside. It conveniently utilizes the first BEOL metal layer closest to the silicon device layer as the reflector to increase directionality. As such, no additional processes will be required for the formation of the reflector. In addition, the gratings are fully-etched, which removes process variations associated with non-uniformity in etching thickness. Inverse design[38] is utilized for the grating design to generate a Gaussian-like diffracted pattern that matches with the fiber mode.

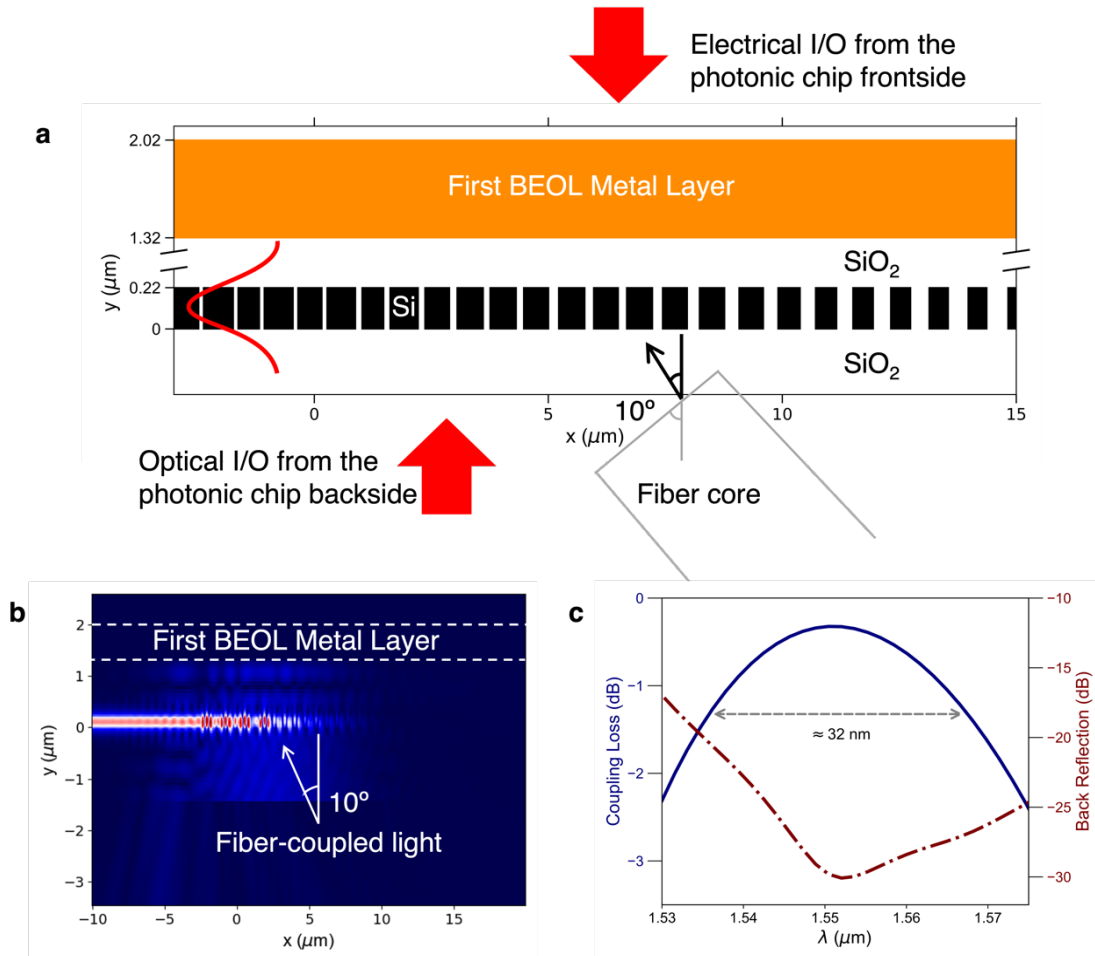

**Figure 4. High-efficiency grating coupler commensurate with standard, wafer-level foundry processes for I/O on the SuMMIT platform**. **a,** The etch profile (Y-X plane cut), obtained via inverse design for mode-matching to the fiber mode, where the first BEOL metal layer (in orange), vertically spaced 1.1 μm away is exploited as the reflector. 220 nm full silicon etch is implemented to avoid variation in grating performance attributed to non-uniformity in etching depths. The direction of coupled-fiber light is illustrated, angled at 10º to the surface normal. The frontside and backside of the PIC wafer are labeled for clarity. **b,** Electric-field distribution of the lightwave in the Y-X plane as it is coupled from the SMF to the waveguide. **c,** Wavelength-dependent coupling loss and back reflection of the grating coupler simulated using 3-D FDTD. 1-dB bandwidth of the grating coupler is 32 nm with a peak coupling efficiency of 93 % (insertion loss = 0.32 dB).

The fully-etched gratings are defined on a 220 nm silicon device layer. The first

BEOL metal layer (0.7 µm-thick copper) is vertically spaced 1.1 µm from the top of the silicon waveguide. In our design, we have constrained the grating coupler to include 25 grating periods. With regards to the first order Bragg condition, the grating period is defined according to equation (1), where λ and $n_{eff.}$ refer to wavelength of coupled light and effective index of each grating period within the coupler, respectively. The linear apodization of the grating coupler can be defined through equation (2), where $DC_i$, $\varphi$, and $x_i$ refer to the localized duty cycle, linear apodization factor and the position along the propagational direction of the lightwave, respectively. $x_0$ refers to the starting point of the first grating period within the model along the X-coordinate. As we have implemented a constraint of 25 periods for subsequent optimization through inverse design, the maximum value of i would be 24.

$$\Lambda = \frac{\lambda}{n_{eff.} - sin\theta} \quad (1)$$

$$DC_i = DC_0 - \varphi(x_{i-1} - x_0) \quad (2)$$

As this work pertains to the design of full etched grating couplers, the effective index of the optical mode in the fully etch regions cannot be evaluated through eigenmode solvers. However, one can circumvent this issue by labeling the effective index difference of the 220 nm silicon and fully etch regions as $\alpha$ ($n_{220\ nm}$ - $n_{etch}$) and the local $n_{eff,i}$ of each grating period in equation (3), where $n_{220}$ and $n_{etch}$ refers to the effective index of the optical mode in the full height silicon (220 nm) and etched regions respectively. With the above, we can rewrite the generalized form of the first order Bragg condition as equation (4) where $DC_0$, $x_0$, $\varphi$, $\beta$ and $\alpha$ are determined analytically prior to inverse design within 2-D finite-difference time-domain (FDTD) where each of the 25 grating periods is optimized independently.

$$n_{eff,i} = n_{etch} + (DC_i \times \alpha) \quad (3)$$

$$\Lambda = \frac{\lambda}{n_{etch} - sin\theta + DC_i\alpha} = \frac{\lambda}{\beta + DC_i\alpha} \quad (4)$$

In Fig. 4a, the optimized grating profile in the Y-X plane subject to the constraint of 100 nm minimum feature size is shown. The constraint is consistent with the typical critical dimension in contemporary silicon photonic foundries. The grating coupler has a length of 18.2 μm, and the electric-field distribution in the Y-X plane as the light is injected from a fiber mode is shown in Fig. 4b. As indicated in Fig. 4c, coupling efficiency as high as 93% (corresponding to an insertion loss of 0.32 dB) can be attained at 1550 nm wavelength with a 1-dB bandwidth (1536 – 1568 nm) of 32 nm. Minimum and maximum back reflection of 20 and 30 dB are achieved within the wavelength region of 1536 – 1568 nm (Fig. 4c).

The grating couplers presented in this section enables scalable implementation of highly-efficient grating couplers within the SuMMIT platform. The feature sizes are compatible with current silicon photonic foundries and thus the device is amenable to wafer-scale processing. These results decouple the electrical and optical I/O's to the frontside and back side of the platform, highlighting the pathway towards high-density optoelectronic packaging.

**Ultra-Efficient Antennas for Beam Steering**

In recent years, Optical Phased Arrays (OPAs) have emerged as a key technology enabled by silicon photonics with widespread applications ranging from Light Detection And Ranging (LiDAR)[8], optical tweezing[39], high-resolution 3D printing[40], free-space projectors[41,42], data communications[43] and more[41]. OPAs are the optical counterpart of the radio-frequency phased arrays which are nowadays commonplace in radar. The reduction in the operating wavelength enables OPAs to achieve diffraction-limited resolution in excess of 1000 times of what is achievable with radio waves, since the angular aperture scales linearly with wavelength.

The basic building block of OPAs is an optical antenna, which is tasked with emitting light from the chip into free space. Accurate control of the position, phase and amplitude of the emitted wave of each individual antenna is critical to creating well-directed beams with high angular resolution[44]. Beam steering is achieved by applying a phase gradient across the antennas, or by tuning the operating wavelength, which shifts the emission angle of the antennas[45].

The beam quality and intensity of the OPA is determined by the array efficiency. In the simplest case, i.e., when each antenna can be approximated as a small dipole of identical scattering characteristics, the array efficiency can be separated into an element factor solely dictated by the properties of an individual antenna and an array factor which accounts for the interactions between the antenna elements. For a series of uniformly spaced antennas with a linear phase gradient and equal emission intensity, the array factor approximately reduces to:

$$\mathcal{R}(\phi) = \frac{\sin^2\left(\frac{N}{2}(\psi+\vec{k}\cdot\vec{a}\cos(\phi))\right)}{\sin^2\left(\frac{1}{2}(\psi+\vec{k}\cdot\vec{a}\cos(\phi))\right)} \quad (5)$$

where $\phi$ gives the emission angle, $\psi$ is the linear phase shift applied between neighboring antennas, $\vec{k}$ represents the wave vector, $\vec{a}$ denotes the array pitch, and N corresponds to the total number of antenna elements. The last two parameters are of critical importance: increasing the number of elements reduces the beam waist and increases resolution, while reducing the spacing between the antennas (ideally $\leq \lambda/2$) minimizes the number of unwanted side-lobes present in the far-field emitted beam.

Various OPA architectures have been experimentally demonstrated, including: 1) a splitter-tree architecture[46], where a series of N 1 × 2 splitters is used to redistribute the input light into $2^N$ rows of antennas, each with their own individual phase shifters; 2) a cascaded architecture[47], where an array of tap couplers is used to extract light from a common bus waveguide along which phase shifters are placed in between the tap couplers; and 3) a 2-D grid architecture[8,48], where each antenna is placed in a 2-D grid and assigned its own individual phase shifter. Although both the splitter-tree and cascaded architectures have advantages related to their simple phase-shifter driving schemes and capabilities to bring the antennas close to each other, their requirement for a tunable laser in order to achieve full 2-D beam steering poses a major limitation to their implementation in real-world scenarios. As a result, a lot of attention has been given to the development of compact and efficient antennas that are compatible with the 2-D OPA architecture[8,48]. For example, compact plasmonic antennas can achieve good directionality, but absorb a significant fraction of light. Conversely, dielectric antennas based on diffractive elements tend to lack sufficient grating strength to achieve high diffraction efficiency and directionality, an issue that is further exacerbated by the size requirements of the antenna. For instance, Sun *et al.*[48] designed an antenna with a total emission efficiency of 86%, but with only 59% directionality. More recently, Khajavi *et al.*[49] addressed this issue by proposing a sub-wavelength metamaterial grating with a simulated diffraction efficiency of 89% and directionality of 94%, which, although compelling, made use of 300 nm thick SOI device layer to increase the grating strength and is thus not compatible with the 220 nm thick SOI used in most photonics foundries.

Similar to the grating couplers proposed in the previous section, we propose a simple antenna design taking advantage of our inverted photonic substrate architecture, where the first BEOL metal layer acts as a reflector. Our proposed device boasts a total diffraction efficiency as high as 97% while still being compact in size, with a length of 7.3 μm and a width of 7.8 μm. The antenna consists of three grating periods, with an etch depth of 130 nm which is used for the definition of silicon slabs in the foundry process[1]. The period and duty cycle are chosen to be 1.068 μm and 0.4 respectively as illustrated in Fig. 5a, obtained via particle swarm optimization in 3-D FDTD. The antenna adopts a focusing structure with a fan angle of 28º, which connects to a SOI waveguide with a width of 0.5 μm. Similar to the grating coupler proposed, the diffracted beam is directed to the PIC wafer backside.

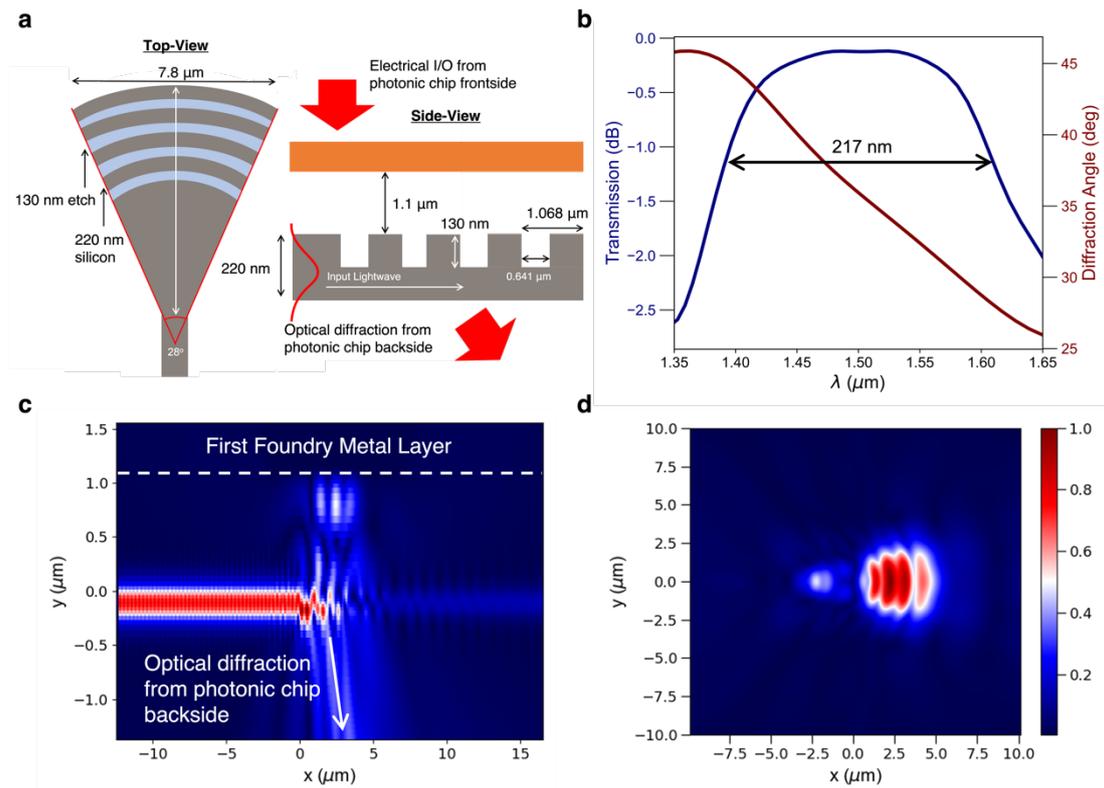

**Figure 5. A high-efficiency antenna for beam steering applications on the SuMMIT platform. a,** Top down and side views of the antenna are illustrated. 130 nm etch depth is used for defining the grating teeth, which is a standard etch depth used in the formation of silicon slabs in photonic foundries. The first BEOL metal layer, spaced 1.1 µm for from the 220 nm silicon device layer in the active PIC stack is used to increase the emission directionality. As illustrated in the side-view schematic, the lightwave propagates along the waveguide and is diffracted to the PIC backside. Only the first BEOL metal layer is illustrated. The frontside and backside of the PIC is indicated. **b,** Coupling loss and diffraction angle against wavelength, showing a 1-dB bandwidth of 217 nm. **c,** Electric-field distribution of the lightwave as it propagates along the waveguide and is diffracted to free space from the PIC backside. **d,** Near-field emission pattern of the antenna.

The simulated diffraction efficiency and angle of the antenna are plotted in Fig. 5b. Peak diffraction efficiency exceeding 97% can be attained with a 1 dB-bandwidth of 217 nm. The simulated electric field of the lightwave as it propagates along the input waveguide and is diffracted from the antenna is displayed in Fig. 5c, and the corresponding near-field optical emission profile is shown in Fig. 5d.

The far-field radiation pattern shown in Fig. 6a in terms of the polar ($\theta$) and azimuthal angles ($\phi$) is derived from the near-field profile. The beam diffracts at 32.76º from the vertical axis ($\theta$), with a full-width half maximum of 13.17º and 30.02º along the $\theta$ and $\phi$-coordinates as indicated in Figs. 6b and 6c respectively.

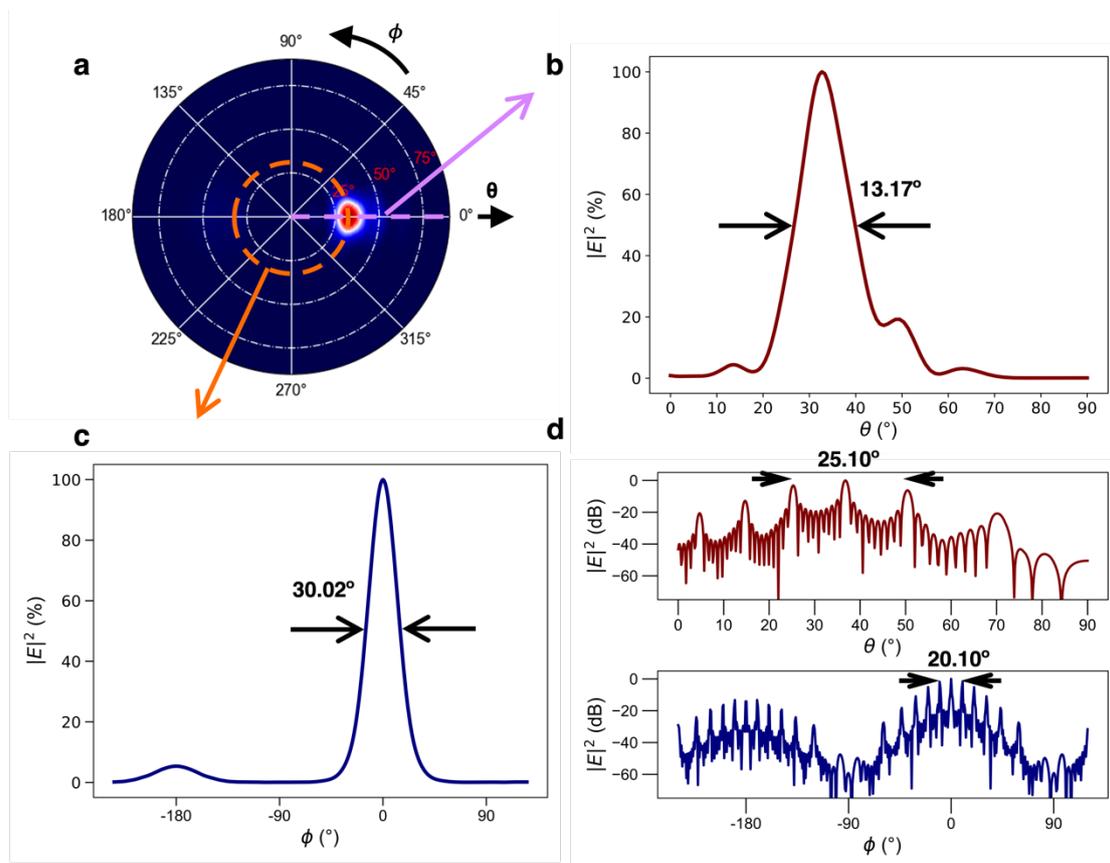

**Figure 6. Far-field characteristics of the high-efficiency antenna. a,** The far-field radiation pattern of the antenna obtained from the near-field emission in Fig. 5c. **b-c,** The emission profiles at **b** $\phi = 0º$ and **c** $\theta = 32.76º$. **d,** Same as **c**, but assuming a 10 × 10 array with a pitch of 9 μm.

To assess the performance of the antenna in a 2-D phased array, we consider an array comprising 10 × 10 antennas a uniform pitch of 9 μm. The far-field emission pattern of the 2-D array is given by the product of the individual antenna far-field and the array factor function. In Fig. 6d, the array emission profile along $\phi = 0º$ is computed in this manner. For the given conditions, the simulated lobe-free steering range of $\theta \times \phi$ is 25.1º × 20.1º.

Similar to the grating couplers presented in the previous section, the antenna design shows that highly efficient components can be realized on the SuMMIT platform. The etching depths as well as requirements on feature sizes are compliant with the process design kits of current photonic foundries, which will facilitate their scalable manufacturing and integration.

**Polarization Rotators enabled by Bi-facial Patterning**

The PIC substrate removal step in Fig. 2 leaves a flat wafer surface without topologies. Therefore, an additional lithographic patterning and etching step can be carried out on the wafer backside. Here we use a polarization rotator as an example to illustrate how this unique bi-facial or double-side patterning capability can be harnessed to create novel photonic device structures.

Of the transverse electric (TE) and transverse magnetic (TM) polarizations, a disproportionate amount of work has been dedicated to the TE polarization in the prevailing 220 nm single-mode SOI waveguide platforms. This is because the output polarization of diode lasers is usually TE, and TE polarization also has stronger lateral confinement, permitting smaller bending radii and higher integration density. However, the TM polarization is useful for several reasons. First, the TM mode is less susceptible to scattering loss incurred by sidewall roughness and in general suffers from lower propagation loss than the TE mode[50]. Moreover, TM mode exhibits enhanced modal overlap with waveguide top cladding. These two considerations combined favor TM mode for sensing applications. The latter factor also implies that TM-mode waveguides could be better suited for certain HI scenarios where larger optical confinement in the deposited or bonded layer overshadows the benefits of using TE mode. Lastly, TM mode is a more convenient choice for on-chip magnetooptical isolators, since breaking out-of-plane structural symmetry is far easier to implement than breaking in-plane symmetry demanded by TE mode isolation[51]. Therefore, an efficient way to perform polarization rotation is valuable in photonic integration.

Thus far, there are several architectures for polarization rotators. These devices can be enabled through cascading several bent sections, where the polarization rotation efficiency is optimized through the radius of curvature[52]. Another approach involves creating a cut edge in a waveguide such that the two lowest order modes are hybridized in the polarization rotation section[53]. Polarization rotation can be facilitated in which the two modes interfere over a beat length of $L_\pi = \pi/{\beta_1 - \beta_2}$ where $\beta_1$ and $\beta_2$ refers to the propagation constant of the two lowest order modes. Yet another scheme capitalizes on mode anti-crossing in asymmetrically cladded waveguides to transition between TM0 and TE1 modes[54–56].

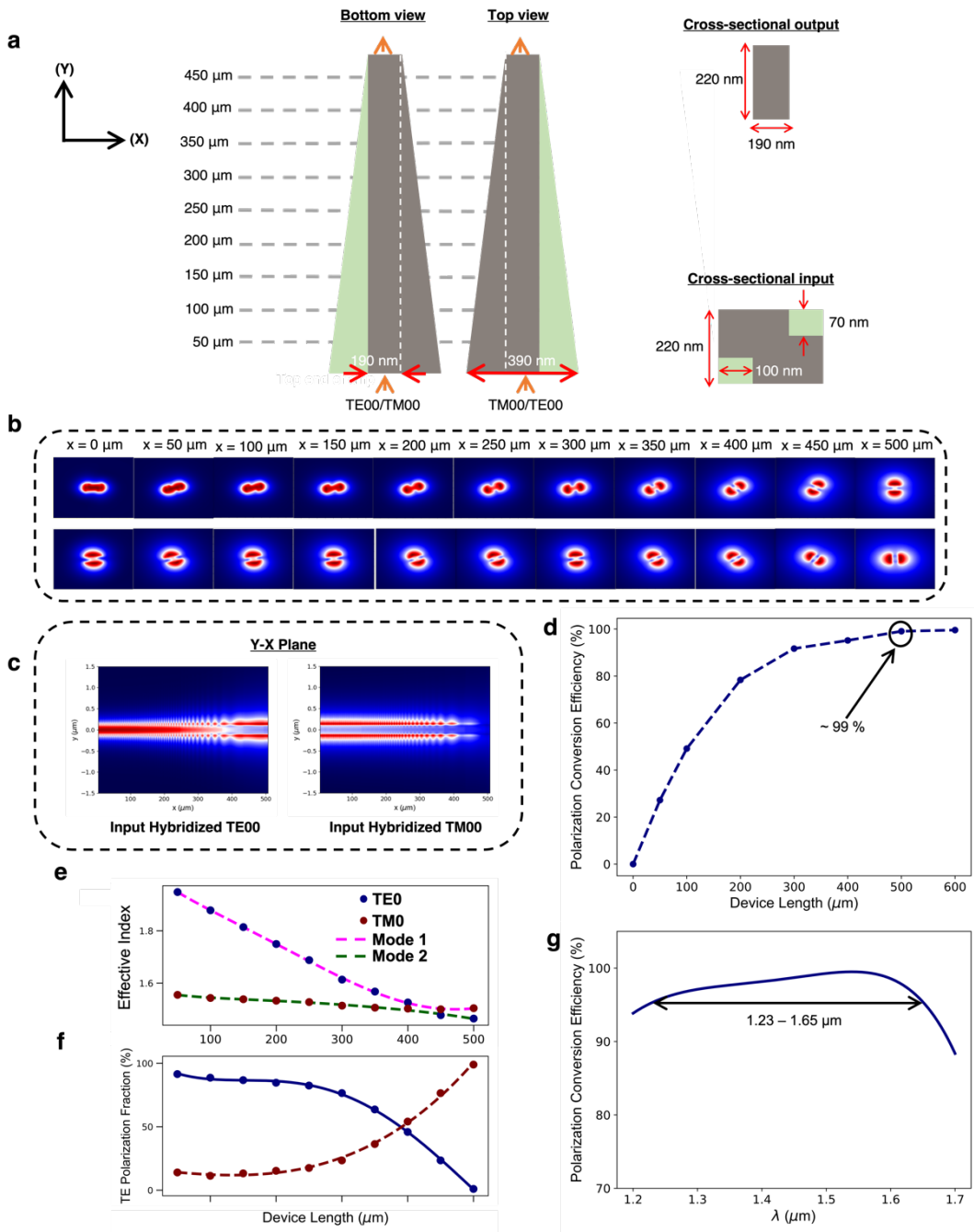

**Figure 7. Adiabatic polarization rotator enabled by double-side waveguide patterning. a,** (Left) illustrations of the polarization rotator in top and bottom views; (right) cross-sectional cuts of the rotator at its input and output. **b,** Cross-sectional electric-field distributions of the hybridized waveguide modes as it travels along the polarization rotator, where the input polarization are hybridized TE0 and TM0 in the top and bottom figures respectively **c,** Electric-field distribution in the Y-X plane as hybridized TE0 (top figure) and TM0 (bottom figure) inputs travel along the rotator and is converted to TM0 and TE0

polarizations, respectively. **d,** Polarization conversion efficiency of the polarization rotator as a function of length, where the optimal length is found to be 500 μm at λ = 1.55 μm. **e,** Effective index and **f,** TE polarization fraction as the hybridized modes propagates along the polarization rotator. **g,** Wavelength dependence of the polarization rotator from λ = 1.2 – 1.7 μm.

The device starts off with a 390 nm-wide asymmetric waveguide with its cross-section shown in Fig. 7a, supporting a hybridized TE0 and TM0 modes; the hybridized modes at the input (Fig. 7b, x = 0 μm) can be attained adiabatically from a 390 nm-wide strip waveguide as described in Supplementary S1. The etch depths at the top and bottom are 70 nm with a width of 100 nm; the top and bottom view of the polarization rotator is show in Fig. 7a. The 70 nm etch layer is widely available in most foundries for the formation of grating couplers[1]. Along the polarization rotation section, the waveguide width is reduced from 390 nm to 190 nm along a length of 500 μm (Fig. 7a). Figure 7b shows the adiabatic evolution of the optical mode along the polarization rotator, where hybridized TE0 transitions to TM0 (top) and hybridized TM0 transforms to TE0 (bottom). At 1.55 μm, polarization conversion efficiency exceeds 99% for a polarization rotator length of 500 μm. The mode evolution of an input hybridized TE0 mode (Y-X plane) as it propagates along the polarization rotator clearly shows its evolution to the TM0 mode (top, Fig. 7c). Conversely, the mode evolution as an input hybridized TM0 converts to TE0 is also presented (bottom, Fig. 7c). The effective indices of the input hybridized TE0/TM0 mode as it propagates along the length of the rotator is shown in Fig. 7e, where an anti-crossing occurs at x = 424 μm; Mode 1 and 2 refers to the modes with the first and second highest effective index. In Fig. 7f, we plot the TE polarization fraction[57] of an input mode along the rotator, affirming the results obtained in Fig. 7d. Note that the TE polarization fractions of the input hybridized TE0/TM0 modes do not equal 0 or unity, which is a result of the adiabatic transition from a 390 nm-wide strip waveguide as discussed in Supplementary S1. The adiabatic nature of the device warrants its broadband operation, and polarization conversion efficiency in excess of 95% can be maintained across the wavelength range of 1.23 to

1.65 μm (Fig. 7g).

The polarization rotator example highlights the 3-D structuring capability enabled by the bi-facial patterning process. Other possibilities include corrugated grating couplers where highly-efficient emission can be attained by designing the gratings in two layers such that the lightwave interferes constructively in the upward direction, and destructively in the downwards direction[35,58].

**Conclusion**

Heterogeneous integration[11,12,16–19] represents the key solution to expand the functional repertoire available to foundry-processed photonic platforms. In this article, we propose the SuMMIT platform, leveraging standard foundry active PIC processes to enable heterogenous silicon photonics. Building on advanced 3-D integration technologies, heterogeneous integration with beyond-CMOS materials can be realized on the wafer scale. The platform is also commensurate with high-density packaging by facilitating bi-facial electrical/optical I/O's. Furthermore, we show that by exploiting the full BEOL stack of the active PIC and double-side patterning capacity, both unique to our platform, a cohort of high-performance photonic devices can be realized. To that effect, grating couplers with efficiency as high as 93%, antennas with diffraction efficiency of 97% and a bandwidth of 217 nm, and a polarization rotator with polarization conversion efficiencies in excess of 99% have been proposed based on foundry-compatible designs. Experimental demonstration is currently underway and far-reaching impacts of the SuMMIT platform on heterogeneous photonic integration will unfold in the near future.

**Acknowledgements**

S.J.X.B. is supported by the Ministry of Education/NTU College of Engineering (CoE) International Postdoctoral Fellowship.

**Author Contributions**

J.H. supervised the project. S.J.X.B., L.R., and J.H. conceived the platform and devices. S.J.X.B., L.R., and D.K.P. performed the simulations. All authors participated in the analysis of data and contributed to the writing of the manuscript.

**Conflict of interest**

The authors declare no competing interests.

**Supplementary information**

Supplementary materials are available at the online version.

# Supplementary Information for Multi-Material heterogeneous integration on a 3-D Photonic CMOS platform

Luigi Ranno[1,†], Jia Xu Brian Sia[1,2,†], Khoi Phuong Dao[1] and Juejun Hu[1,*]

[1]Department of Materials Science & Engineering, Massachusetts Institute of Technology, Cambridge, M.A., USA.

[2]Centre for Micro- & Nano-Electronics (CMNE), Nanyang Technological University, Singapore

[†] These authors contributed equally to this work.

Correspondence: Juejun Hu

Email: hujuejun@mit.edu

# Adiabatic conversion of a strip waveguide to the bi-facial asymmetrical waveguide for polarization rotation

The adiabatic converter illustrated in Fig. S1a will enable the adiabatic conversion between a 390 nm-wide strip waveguide to the asymmetrical waveguide implemented at the input of the polarization rotator. The top, bottom as well as the cross-sectional illustration (input and output) of the device is shown in Fig. S1a. In Fig. S1b, the evolution of the input TE0 (top) and TM0 (bottom) optical mode into hybridized modes as it propagates along the adiabatic converter is shown in Fig. S1b. The TE polarization fraction of the input TE and TM modes along the length of the converter is illustrated in Fig. S1c, where input TE0 optical mode with TE polarization fraction of ~ 96% is converted into a hybridized mode with TE polarization fraction of 86%. Likewise, the conclusion applies for an input TM polarization.

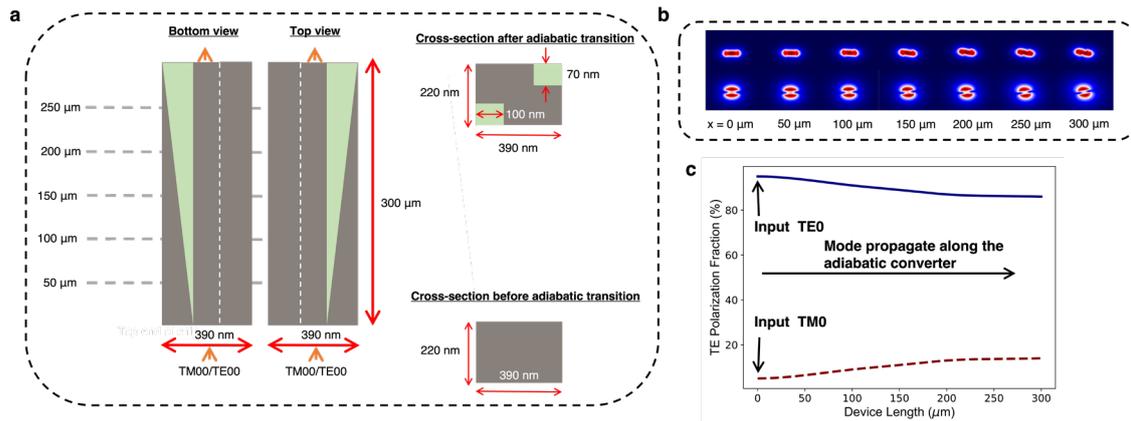

**Figure S1. Adiabatic converter of a 390 nm-wide strip waveguide to the asymmetrical waveguide at the input of the polarization rotator (Fig. 7a). a,** (Left) illustrations of the polarization rotator in top and bottom views; (right) cross-sectional cuts of the adiabatic converter at its input and output. **b,** Cross-sectional electric-field distributions of the hybridized waveguide modes as it travels along the adiabatic converter, where the input polarization are TE0 and TM0 in the top and bottom figures respectively **c,** TE polarization fraction with regards to input polarization TE0 and TM0 (At Device Length = 0 μm).